\documentclass[preprint]{aastex701}
\usepackage{amsmath}
\usepackage{graphicx}
\usepackage{soul}
\usepackage{xcolor}
\usepackage[normalem]{ulem}
\usepackage{cancel}
\usepackage{lineno}

\definecolor{rust}{rgb}{0.72,0.25,0.05}
\newcommand{\Df}{D_{\mathrm{f}}}

\begin{document}

\title{Current Sheet Fragmentation in the Course of Repeated Eruptive Magnetic Flux Emergence}

\author[orcid={0009-0008-6993-203X}]{Vaggelis Karantanis}
\email{e.karantanis@uoi.gr}
\affiliation{Department of Physics, University of Ioannina, 45110 Ioannina, Greece}

\author[orcid={0000-0002-8700-4172}]{Loukas Vlahos}
\email{loukasvlahos@gmail.com}
\affiliation{Department of Physics, University of Ioannina, 45110 Ioannina, Greece}
\affiliation{Department of Physics, Aristotle University, 54124 Thessaloniki, Greece}

\author[orcid={0000-0001-9782-2294}]{Heinz Isliker}
\email{loukasvlahos@gmail.com}
\affiliation{Department of Physics, University of Ioannina, 45110 Ioannina, Greece}
\affiliation{Department of Physics, Aristotle University, 54124 Thessaloniki, Greece}

\author[orcid={}]{Angelos Giannis}
\email{a.giannis@uoi.gr}
\affiliation{Department of Physics, University of Ioannina, 45110 Ioannina, Greece}

\author[orcid={}]{Vasilis Archontis}
\email{archontis@uoi.gr}
\affiliation{Department of Physics, University of Ioannina, 45110 Ioannina, Greece}

\begin{abstract}
Magnetic reconnection between emerging magnetic flux and the ambient coronal field fundamentally drives solar jets. We use three-dimensional resistive-MHD simulations of eruptive flux emergence to investigate how the large-scale current sheet formed at the interface fragments during recurrent jet activity. The simulations show that jet formation and current-sheet disruption are coupled elements of a single multiscale process. The current structures developing along open field lines and within the jet are relatively weak, exhibiting a predominantly filamentary morphology. In contrast, the larger structures linked to the closed-field topology are noticeably stronger, and the eruptive development of the emerging flux further increases both their number and strength. The parallel-current distributions within the closed magnetic field lines region exhibit Gaussian central cores with pronounced non-Gaussian tails, implying a small subset of strong, intermittent structures.  Magnetic spectra follow a power law with a Kolmogorov-like slope in the large-scale range, becoming much steeper in the small-scale range—consistent with reconnection-mediated magnetic structuring. Box-counting analysis shows that the strongest current structures have a fractal dimension $\Df\simeq2$, indicating a fragmented population of sheet-like coherent structures. Cluster analysis further reveals broad, power-law distributions of cluster volumes and their Ohmic dissipation energies, with individual events reaching $\sim10^{20}$--$10^{25}\,{\rm erg}$ over their estimated lifetimes. These results demonstrate that eruptive flux emergence can self-consistently transform a global reconnecting current sheet into a localized, intermittent, sheet-dominated dissipative network, providing a natural environment for impulsive coronal heating and motivating future particle-acceleration studies.
\end{abstract}

\section{Introduction}

The emergence of magnetic flux is among the most effective mechanisms for transporting free magnetic energy and helicity from the solar interior into the corona. As the newly emerged magnetic field expands into the already magnetized atmosphere, it encounters the pre-existing ambient field and naturally forms an extended current sheet at their interface \citep{Heyvaerts77}. Magnetic reconnection across this current layer alters field connectivity, heats and accelerates the plasma particles, and generates collimated outflows, establishing the fundamental scenario for solar jet formation \citep{YokoyamaShibata1996,Archontis2004,Galsgaard2005,Archontis2005,Raouafi2016}.

Three-dimensional simulations indicate that jets produced by flux emergence do not result from a single, smooth reconnection episode. Continued emergence, combined with injected shear and twist, can sustain the reconnecting current layer, create a strongly stressed magnetic core, and drive the system from relatively narrow, standard jets to broader, more explosive blowout jets \citep{Moore2010,ArchontisHood2013,MorenoInsertisGalsgaard2013}. Within this picture, jet formation is intrinsically multiscale: the large-scale expansion of the emerging magnetic structure controls the development of the extended current sheet, whereas the internal dynamics within that sheet determine the outflow’s morphology, intermittency, and energy release.

For some time, it has been recognized that the primary current sheet is unlikely to remain a single, coherent structure \citep{Onofri06, Guo21,  WangYulei25}. In plasmas with very high Lundquist numbers, extended current sheets are prone to the tearing instability, which causes them to break up into plasmoids or small, flux-rope-like structures and, within a few hundred Alfvén times, to transition into an MHD turbulent state. This evolution produces intermittent reconnection and a hierarchical spectrum of energy-release events \citep{ShibataTanuma2001,Loureiro2007,Wyper2016,Kumar2019}. Consequently, the large-scale current layer that powers the jet is naturally linked to the localized brightenings, blobs, secondary islands, and fine internal strands that are increasingly observed in eruptive jets. The kinetic study by \citet{Isliker2019} provides additional support for this scenario, demonstrating that the breakup of the large-scale current sheet in an emergence-driven jet generates a highly turbulent, fractal electric-field environment that can efficiently heat and accelerate particles. Their findings strengthen the interpretation that the association of jets with hard X-ray emission, type III radio bursts, and impulsive solar energetic particles originates within this fragmented reconnection region, rather than in a single, smooth acceleration site \citep{Isliker2019,Raouafi2016}.

The core issue, then, is not merely how reconnection initiates a jet, but how the large-scale current sheet that develops during the eruptive interaction between emerging and pre-existing magnetic fields evolves, breaks up, and seeds the fine structures embedded within the jet. Demonstrating this linkage is crucial for a physically consistent description of emergence-driven jets as multiscale reconnection phenomena, in which global magnetic reconfiguration and current-sheet fragmentation represent two interconnected facets of the same dynamical process.

What is still not well understood are the precise links between these different scales: how the large-scale eruptive dynamics generate the primary current sheet; at what point that sheet first becomes vulnerable to tearing instability; how its subsequent fragmentation shapes the jet spire, base brightenings, and fine filamentary patterns; and whether the observed small-scale features are better characterized as plasmoids, secondary flux ropes, or more general three-dimensional, magnetically dominated, turbulent reconnection fragments.

In this article, we present a rigorous demonstration that the initiation of eruptive jets and the disruption of current sheets are intrinsically linked expressions of a single multiscale mechanism. Section \ref{setup} outlines the physical setup together with the numerical techniques used in our analysis. In Section \ref{currentsheetfragmentation}, we detail the time evolution and subsequent fragmentation of the large-scale current sheet that develops during the recurrent eruptions. Section \ref{statistics} is devoted to a statistical examination of the current filaments and their corresponding magnetic fields, including histogram analyses, magnetic energy spectra, fractal spatial organization, and the identification and characterization of clusters formed by fragmented currents. Finally, in Section \ref{summary}, we recapitulate and integrate our principal conclusions.

\section{Numerical Set Up}\label{setup}

We employed the Lare3D code described in \citet{arber2001staggered} to numerically integrate the three-dimensional, time-dependent resistive MHD equations in Cartesian geometry. The MHD equations are cast in dimensionless form and solved numerically, with the primary variables being the density $ \mathrm{\rho} $, velocity $ \mathrm{u} $, pressure $ \mathrm{p} $, magnetic field $ \mathrm{B} $, constant gravitational acceleration, and current density $ \mathrm{J} $, defined via Amp\`ere’s law $ \mathrm{J} = \mathrm{\nabla} \times \mathrm{B} $. We also evolve the specific energy density $ \mathrm{\epsilon} $ and include a uniform resistivity $ \mathrm{\eta} = 0.01 $ in dimensionless units. A fully ionized plasma is assumed, and we account for both viscous heating and Joule dissipation. 

The code integrates the equations numerically in their Lagrangian formulation and subsequently projects the solution back onto an Eulerian grid. The computational domain consists of 420 $\times$ 420 $\times$ 420 grid points along each direction, corresponding to a physical domain of 64.6 $\times$ 64.6 $\times$ 64.6 Mm in the $x$, $y$, and $z$ directions.

We impose periodic boundary conditions along the $y$-direction, and open boundary conditions along the $x$-direction and at the top of the computational domain. The bottom boundary is taken to be closed. The background medium is composed of an adiabatically stratified convection zone for $-4.8 \leq z < 0$ Mm, an isothermal photosphere for $0 \leq z < 1.8$ Mm, a chromosphere/transition region for $1.8 \leq z < 3.2$ Mm, and an isothermal corona for $3.2 \leq z < 59.8$ Mm. The initial profiles of temperature, pressure, and density are the same as in \citet{zhuleku2025recurrent} and \citet{karantanis2026comparative}, which are similar to those in the Sun. 

We also prescribe an ambient coronal magnetic field, directed upward and with a strength of $B = 10$ G, which extends through the underlying convection zone, forming an angle $\theta = 11^{\circ}$ with the $z$-axis and $\phi = 183^{\circ}$ with the $y$-axis. With this configuration, the ambient and emerging magnetic field lines are antiparallel, promoting magnetic reconnection when they interact \citep{galsgaard2007effect}. We monitor the system's evolution for 144 minutes of physical time.

\section{Dynamic evolution of magnetic flux emergence and current sheet fragmentation}\label{currentsheetfragmentation}

\subsection{Evolution of current sheets in the open field region}

In the previously described stratified medium, we introduce a toroidal magnetic flux tube into the convection zone, fixing its footpoints at the lower boundary of the computational domain at $z=-4.8\,\mathrm{Mm}$. The toroidal flux tube model used here follows the approach of \cite{hood2009emergence} and has been extensively described in \citet{zhuleku2025recurrent,karantanis2026comparative}. The maximum strength of the tube's magnetic field is $B_{0} = 6300$ G, dropping exponentially in the radial direction, while the tube's radius is $r_{0}=0.45$ Mm. The twist parameter representing the number of rotations of the field lines around the axis per unit length is set at $\alpha = 0.4$. Since the tube is initially in thermodynamic equilibrium with the ambient plasma, its buoyant ascent is initiated by imposing a density deficit along its central portion, following the configuration used in earlier flux-emergence simulations \citep{Syntelis2017,Chouliaras2023,zhuleku2025recurrent,karantanis2026comparative}.
\begin{figure}[htp!]
    \centering
    \includegraphics[width=0.45\textwidth]{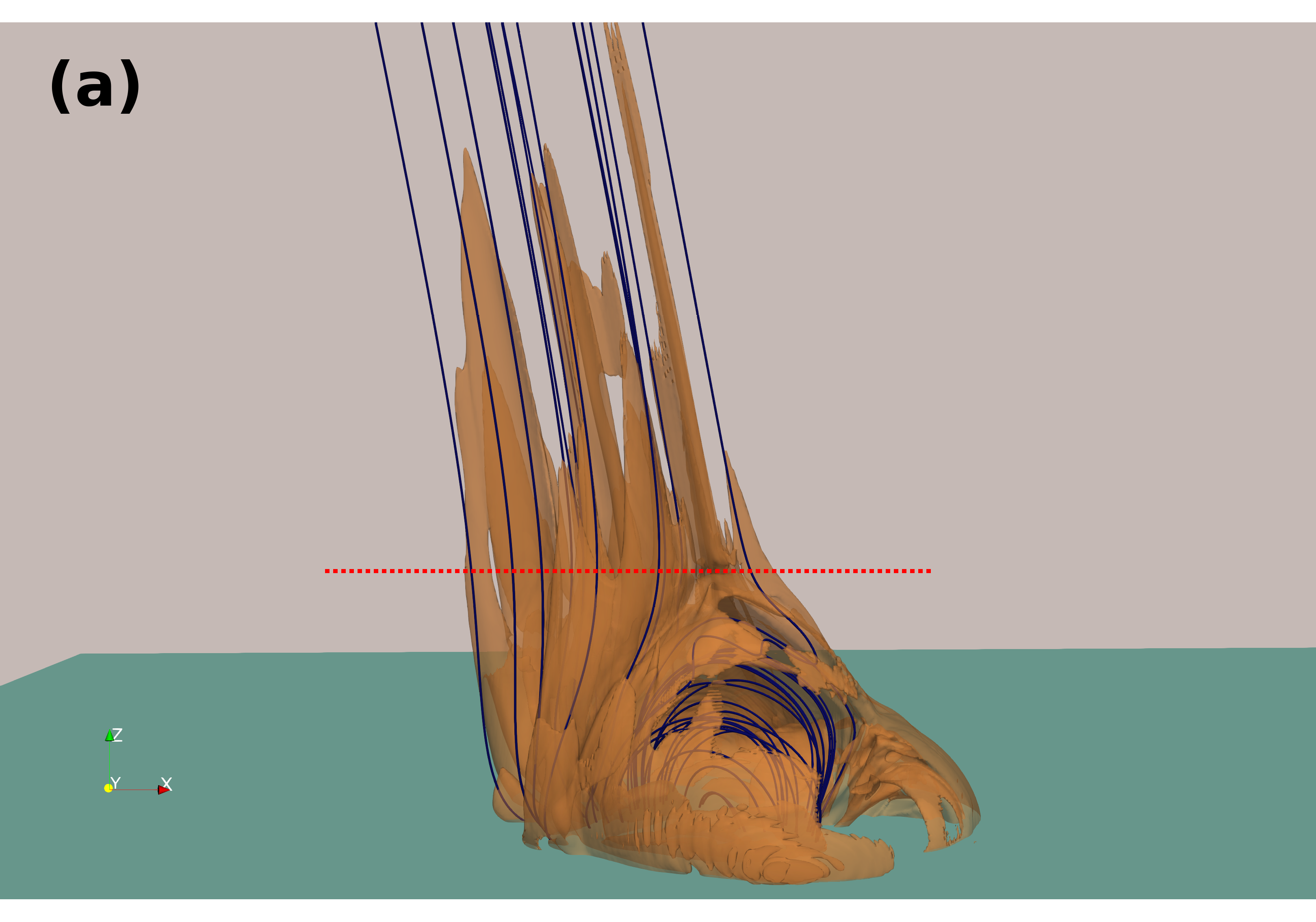}
    \includegraphics[width=0.45\textwidth]{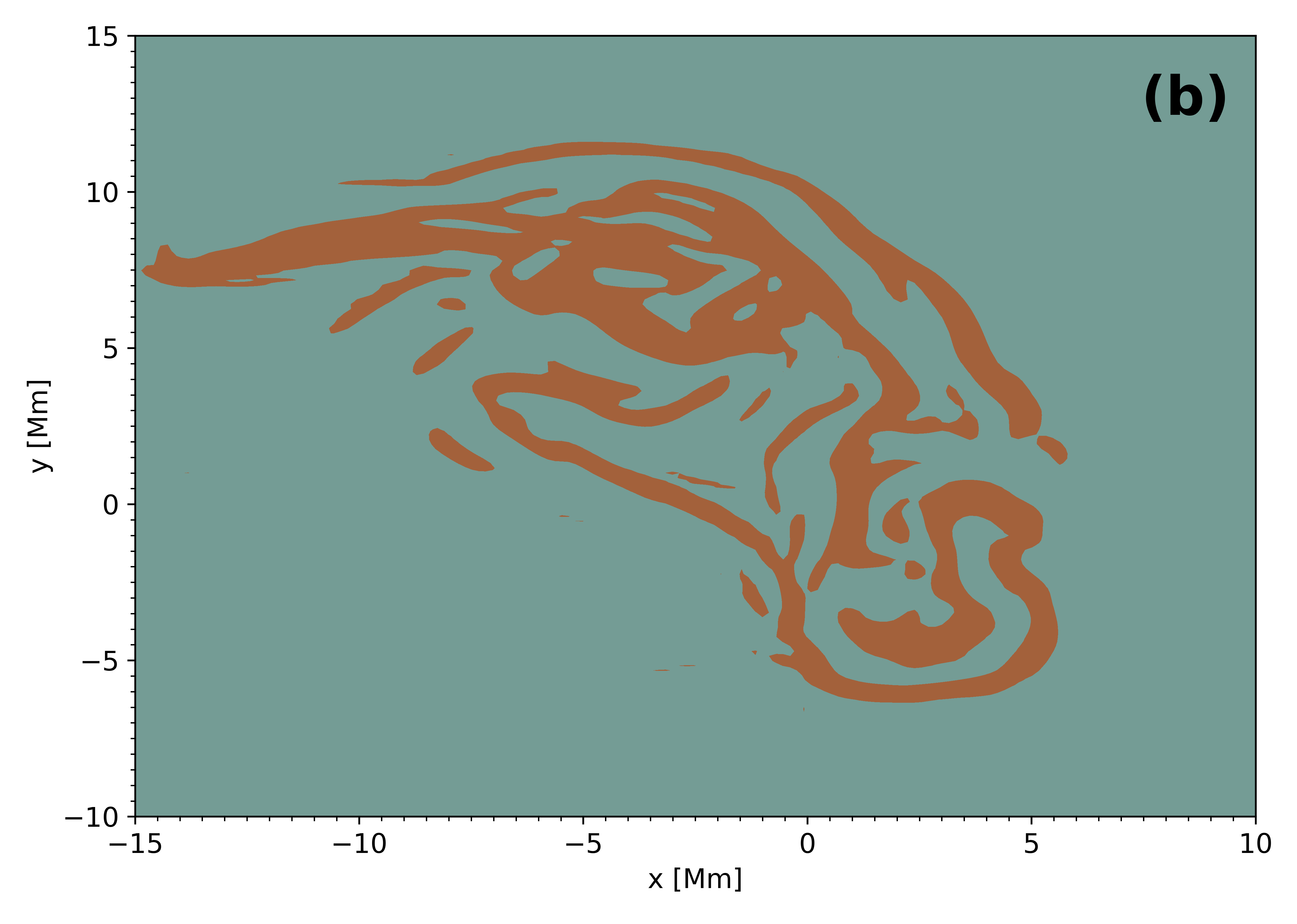}
    \caption{Panel (a) shows the isosurface of the parallel current $J_{\parallel} = 0.0001$ [A m$^{-2}$] at $\mathrm{t}=50$ min. In panel (b), we present the jet's spire in 2D, viewed from the top, at the height marked by the red dashed line in panel (a), $z=21.4$ Mm above the photosphere, showing the same isosurface.}
    \label{fig:J_par_3D_jet}
\end{figure}
Once the density perturbation is introduced, the buoyant flux tube becomes magnetically unstable, prompting it to rise through the photosphere and fan out into the overlying atmosphere. As the emerging magnetic structure reaches the lower corona, it encounters the pre-existing coronal magnetic field. Since the two magnetic systems are oppositely directed, a sharp interface forms between them. As emergence continues, the expanding tube is pushed against the surrounding field, eventually producing an extended current sheet. Magnetic reconnection within this sheet drives bidirectional outflows, ejecting plasma almost vertically in opposite directions. The upward-directed branch of this reconnection outflow forms the so-called standard jet \citep{YokoyamaShibata1996}. The inverse "Y" topology can be distinguished in the complicated 3D configuration in Figure \ref{fig:J_par_3D_jet}(a). The open field lines contain current filaments with low values of $J_{\parallel}$. At the height $z=21.4$ Mm above the photosphere, marked by the red dashed line, Figure \ref{fig:J_par_3D_jet}(b) gives the 2D representation of the current structures in the jet region. This shows in more detail the small parallel currents in the jet's outflow region. Because the open field lines associated with the jet have weaker magnetic fields, the resulting currents are smaller compared with those generated by the flux tube system.

\subsection{Current sheet fragmentation in the closed field lines region}

\begin{figure}[htp!]
    \centering
    \includegraphics[width=\textwidth]{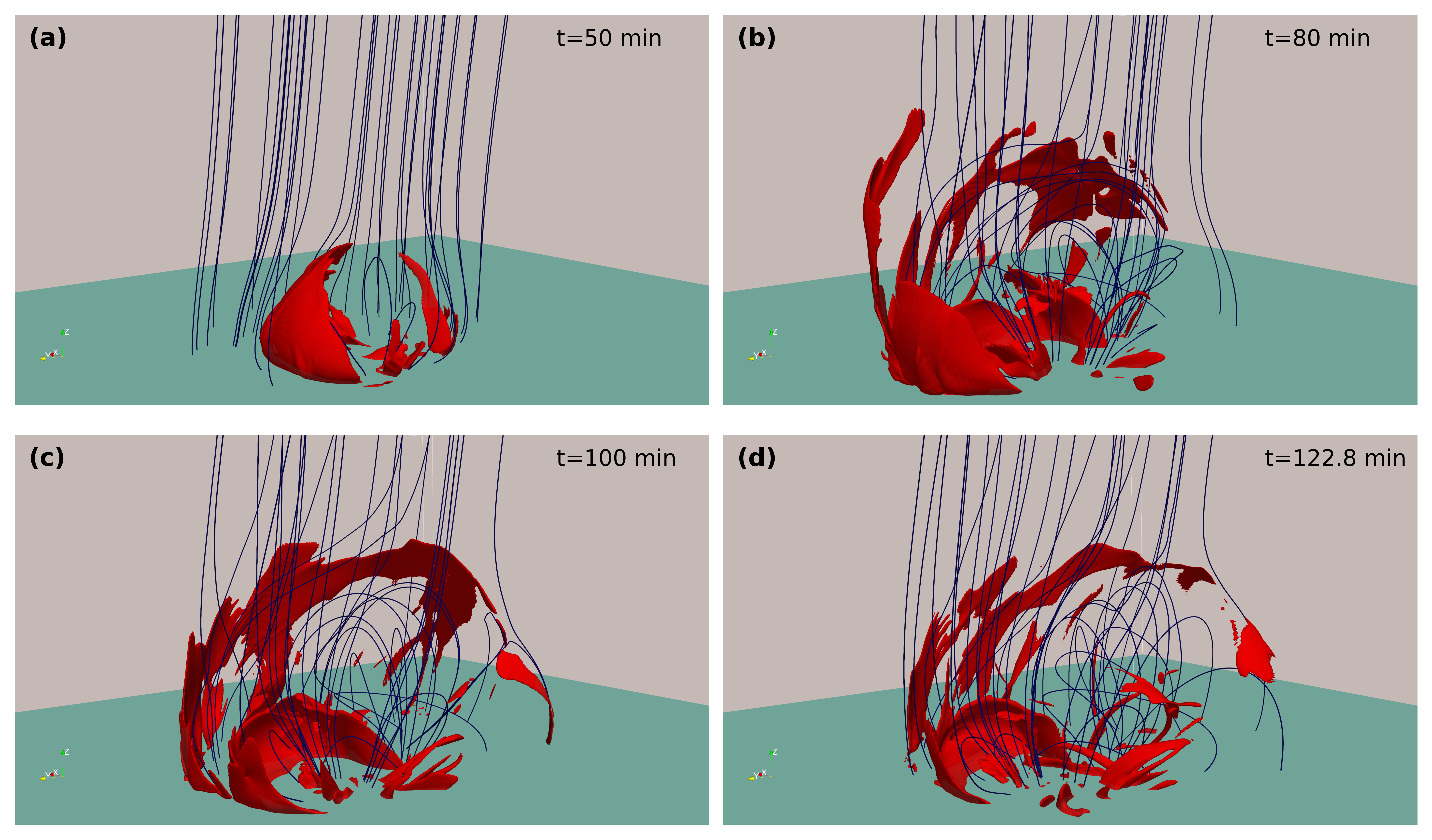}
    \caption{We show an isosurface of the parallel current $J_{\parallel} = 0.002$ [A m$^{-2}$] at $\mathrm{t}=50$ min, $\mathrm{t}=80$ min, $\mathrm{t}=100$ min, and $\mathrm{t}=122.8$ min, in panels (a), (b), (c) and (d) respectively. The magnetic
    field streamlines reveal that the eruptive region is populated by numerous spatially coherent, filamentary structures. An \href{https://drive.google.com/file/d/1h0LMuQUh-DP2JyM_a1iryPuicRMYVJ5B/view?usp=sharing}{\color{red}animation} of this figure is available. The video shows the continuous time evolution of the eruptive region and filamentary structures from $\mathrm{t}=50$ to $\mathrm{t}=144$ min of the simulation.}
    \label{fig:J_par_3D}
\end{figure}

Meanwhile, pronounced shear flows and internal reconnection along the polarity inversion line (PIL) give rise to a new flux rope at comparatively low atmospheric heights. This mechanism has already been reported in previous flux emergence simulations \citep{ArchontisTorok2008,Syntelis2017}. After it forms, the new flux rope ascends and reconnects with the pre-existing coronal magnetic field, initiating an eruption. The eruption propels a blowout jet through the surrounding plasma. In doing so, it further disrupts the magnetic configuration while largely preserving the sheet-like appearance of the coherent structures. Once the plasma from the blowout jet has been expelled, the system settles into a more fragmented configuration, containing more numerous localized structures than it did before the eruption. See panel (b) of Figure \ref{fig:J_par_3D}.

During this stage of evolution, plasma continues to circulate within the original arcade, and the large-scale magnetic configuration becomes fully disrupted. However, continued shearing at the photosphere steadily drives reconnection along the PIL, producing a succession of newly formed flux ropes. Each of these ropes erupts into an already disturbed, finely structured environment, triggering a series of blowouts. Over the 144-minute duration of our simulation, we detect four such blowout jets. As a result, coherent structures build up from one eruption to the next while still preserving their sheet-like character, as shown in panels (c) and (d) of Figure \ref{fig:J_par_3D}. A more detailed analysis of this recurrent eruptive behavior is provided in \citet{zhuleku2025recurrent} and \citet{karantanis2026comparative}.
Figure \ref{fig:J_par_3D}(a) presents the three-dimensional spatial distribution of the parallel current, using an isosurface of $J_{\parallel} = 0.002$ [A m$^{-2}$]. This is at $\mathrm{t} $= 50 min, as the standard jet has formed. Panels (b), (c) and (d) have the same isosurface of the stronger parallel current. They correspond to time $\mathrm{t} $=80 min after the first and second blowout jet, $\mathrm{t} $=100 min after the third blowout jet, and a more settled phase at $\mathrm{t} $= 122.8 min after the fourth and final blowout jet. The isosurface value for $J_{\parallel}= 0.002$ lies in the far tail of the distribution of the fragmented currents, as we will show in the following section. In the same figure, a collection of magnetic field streamlines reveal a highly tangled magnetic configuration, indicating that regions of intense $J_{\parallel}$ are embedded within a complex magnetic environment. A more detailed presentation of the evolution of the strongest $J_{\parallel}$ isosurface and the magnetic configuration can be seen as an \href{https://drive.google.com/file/d/1h0LMuQUh-DP2JyM_a1iryPuicRMYVJ5B/view?usp=sharing}{\color{red}animation}, where the main current sheet fragments into smaller structures.

The high-intensity current-sheet structures produced by the fragmentation of the large-scale current sheet are predominantly concentrated at the boundary between open and closed magnetic field lines. As the eruptive dynamics advect these current sheets upward, they are conveyed along the open-field topology. In contrast, smaller-scale current sheets, such as those shown in Figure \ref{fig:J_par_3D_jet}, occur both within the closed-field domain and within the jet spire, whereby the former continuously supplies plasma to the open-field region via magnetic reconnection between the two magnetic systems.
In the remainder of this article, we focus our analysis on the evolution of the current structures within the closed magnetic field configurations.

\section{Statistical investigation of current sheets throughout their dynamic evolution}\label{statistics}

\subsection{Histogram of the parallel currents and electric field}

We analyze the distribution of the parallel current, $J_{\parallel}$, because we expect the associated current sheets to be aligned with the magnetic field. The results reported in this article are mainly based on the final, more fragmented state of the eruption processes.

\begin{figure}[!hpt]
    \centering
    \includegraphics[width=0.45\textwidth]{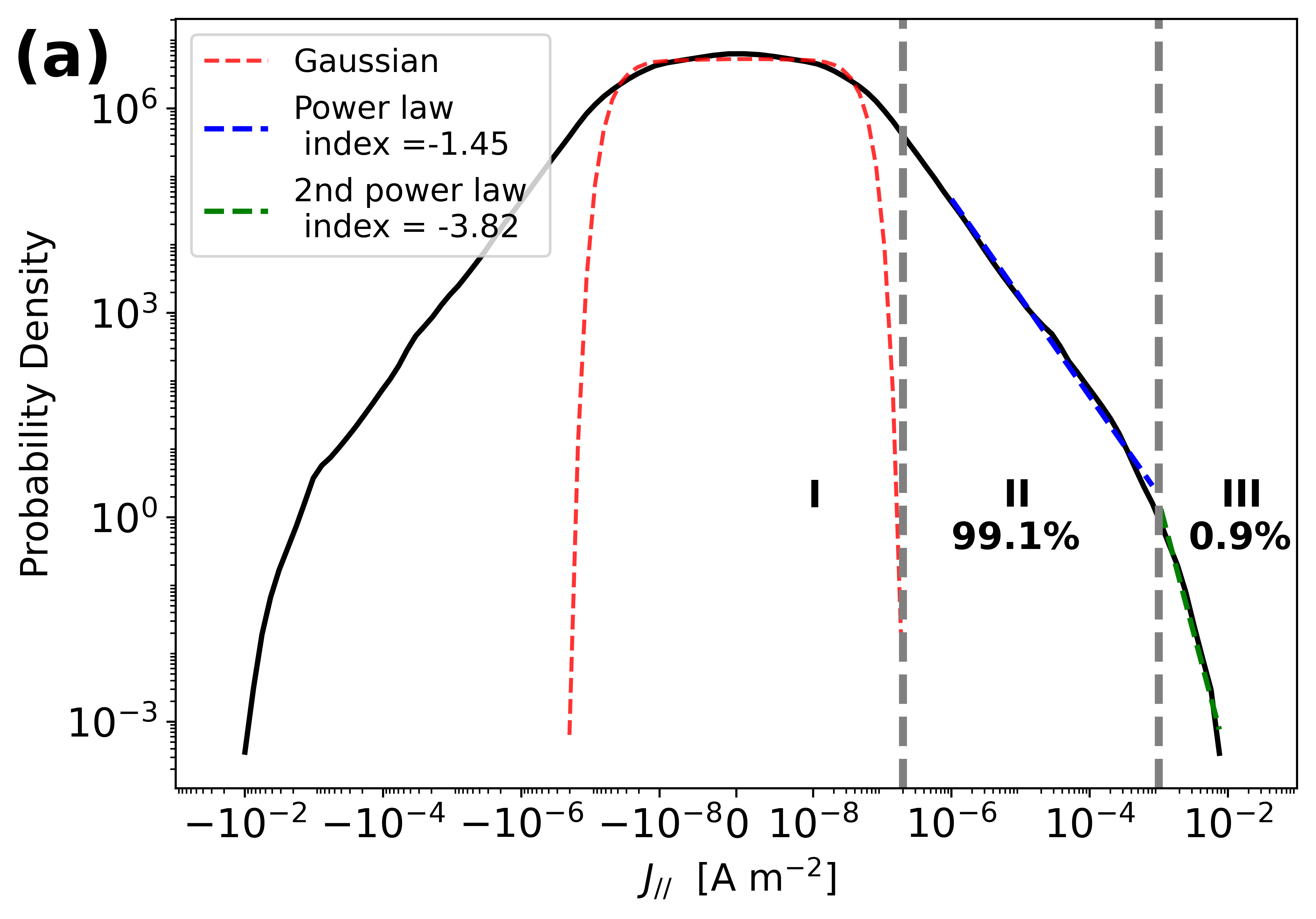}
    \includegraphics[width=0.45\textwidth]{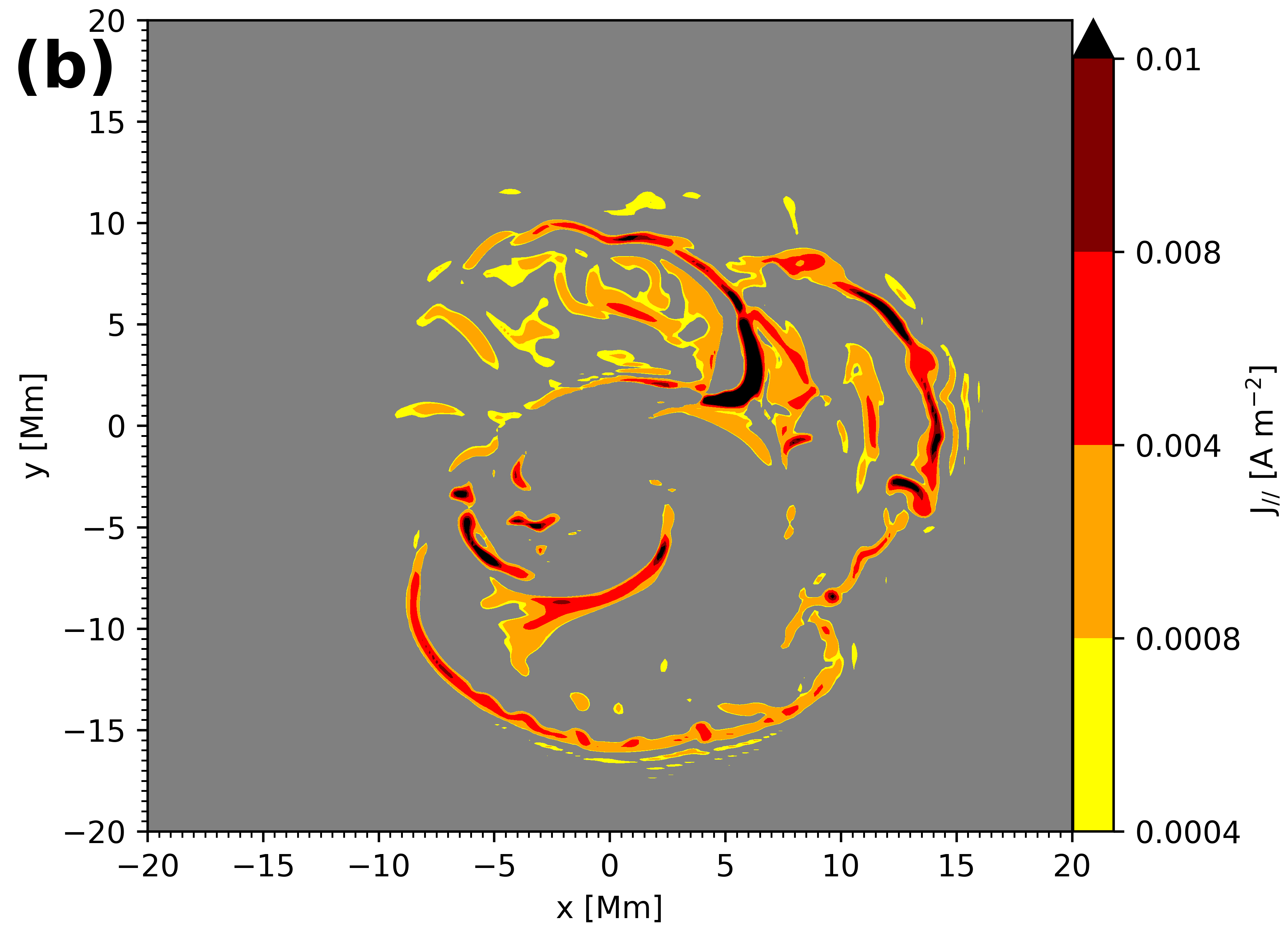}
    \caption{Panel (a) shows the distribution of $J_{\parallel}$ in the corona. A Gaussian region ($\mathbf{I}$) near zero, while the first power law tail ($\mathbf{II}$) has an index of $\alpha=-1.45$ and the second power law tail ($\mathbf{III}$) at bigger values has an index of $\alpha=-3.82$. Panel (b): Top view at an $\mathrm{x}-\mathrm{y}$ cut at $\mathrm{z}=4.47$ Mm above the base of the photosphere, showing several isosurfaces of strong $J_{\parallel}$ from region ($\mathbf{III}$). An \href{https://drive.google.com/file/d/1cw64f8_h3MniN5QRH1cG54WpxLrSBbnu/view?usp=sharing}{\color{red}animation} is available that shows the full evolution with many current sheets appearing and dissipating throughout the whole simulation.}
    \label{fig:J_par_histogram}
\end{figure}

As shown in Figure \ref{fig:J_par_histogram}(a), for a time during the later stages of the simulation (t=122.8 min), the small-amplitude variations with $- 2\times 10^{-7} \leq J_{\parallel} \leq 2\times 10^{-7}\, [\mathrm{A\,m^{-2}}]$  are well represented by a Gaussian, whereas larger magnitudes produce non-Gaussian tails that clearly exhibit power-law behavior. We identify three distinct regions in this histogram: ($\mathbf{I}$) a Gaussian region centered around zero; ($\mathbf{II}$) a range $2\times 10^{-7} < J_{\parallel} < 8\times 10^{-4}\, [\mathrm{A\,m^{-2}}]$ characterized by a power-law tail; and ($\mathbf{III}$) a regime with an even steeper power-law tail for very large currents, $J_{\parallel} > 8\times10^{-4}\, [\mathrm{A\,m^{-2}}]$. This three-part structure remains throughout the simulation.

The distribution spans the full range of positive and negative values and is approximately symmetric around zero.  This characteristic is already present during the formation of the standard jet and persists through all subsequent eruptions, with only slight variations in the corresponding power-law indices. In the histogram of $J_{\parallel}$ at $t=122.8$ min in Figure \ref{fig:J_par_histogram}(a), the first tail (\textbf{II}), immediately beyond the Gaussian core, follows a power law with an index $\alpha=-1.45$. At higher values, the distribution steepens and forms a second power-law tail (\textbf{III}) with an index $\alpha=-3.82$. The heavy non-Gaussian tail indicates that rare, spatially localized coherent structures occur much more frequently than expected under Gaussian fluctuations, and they may dominate dissipation, heating, reconnection, and particle acceleration.

We also determined the fraction of $J_{\parallel} > 0$ values residing in each part of the distribution. Of the total $J_{\parallel} > 0$ values, 60\% lie in the Gaussian core, 39.6\% in the first power-law tail, and 0.4\% in the second, steepest tail containing the strongest parallel currents. If we restrict our attention to the two power-law regimes and exclude the small currents in the Gaussian core, then 99\% of the considered $J_{\parallel}$ values fall within the first power-law tail and almost 1\% within the second tail containing the most intense currents (see Figure \ref{fig:J_par_histogram}a). Because the distribution is fully symmetric around zero, the same percentages apply when $J_{\parallel} < 0$ values are included. The percentage of the very strong current structures in region (\textbf{III}) that possibly reconnect is thus very small, and similar results have been reported for the solar wind by \cite{Hou21}.

In Figure \ref{fig:J_par_histogram}(b), we present a top-down view of the computational domain in the $x$–$y$ plane at a height of $z = 4.47$ Mm, corresponding to low coronal altitudes. The figure shows isosurfaces of $J_{\parallel}$ for various strong current values. In particular, we plot only those values belonging to region ($\mathbf{III}$) of the histogram in Figure \ref{fig:J_par_histogram}(a), where the second power-law distribution emerges. At $t = 122.8$ min, this plot reveals the sheet-like structure of the current sheets even in the later stages of the simulation after four blowout jets. An \href{https://drive.google.com/file/d/1cw64f8_h3MniN5QRH1cG54WpxLrSBbnu/view?usp=sharing}{\color{red}animation} showing the full temporal evolution of the configuration in Figure \ref{fig:J_par_histogram}(b) is available and illustrates the continuous formation and dissipation of thin current sheets.

\subsection{Magnetic field perturbation - Spectral analysis}

\begin{figure}[htp!]
    \centering
    \includegraphics[width=0.45\textwidth]{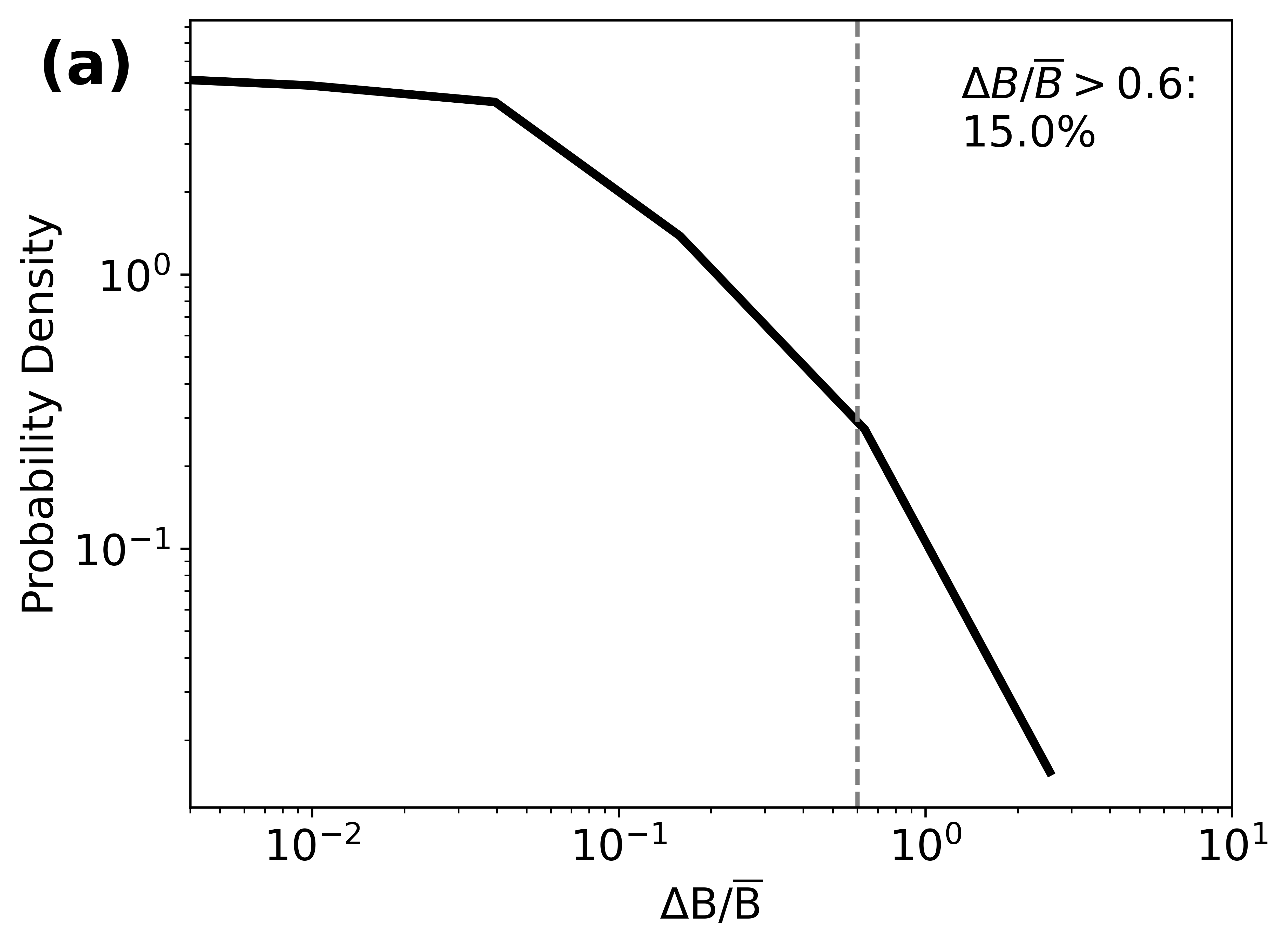}
    \includegraphics[width=0.45\textwidth]{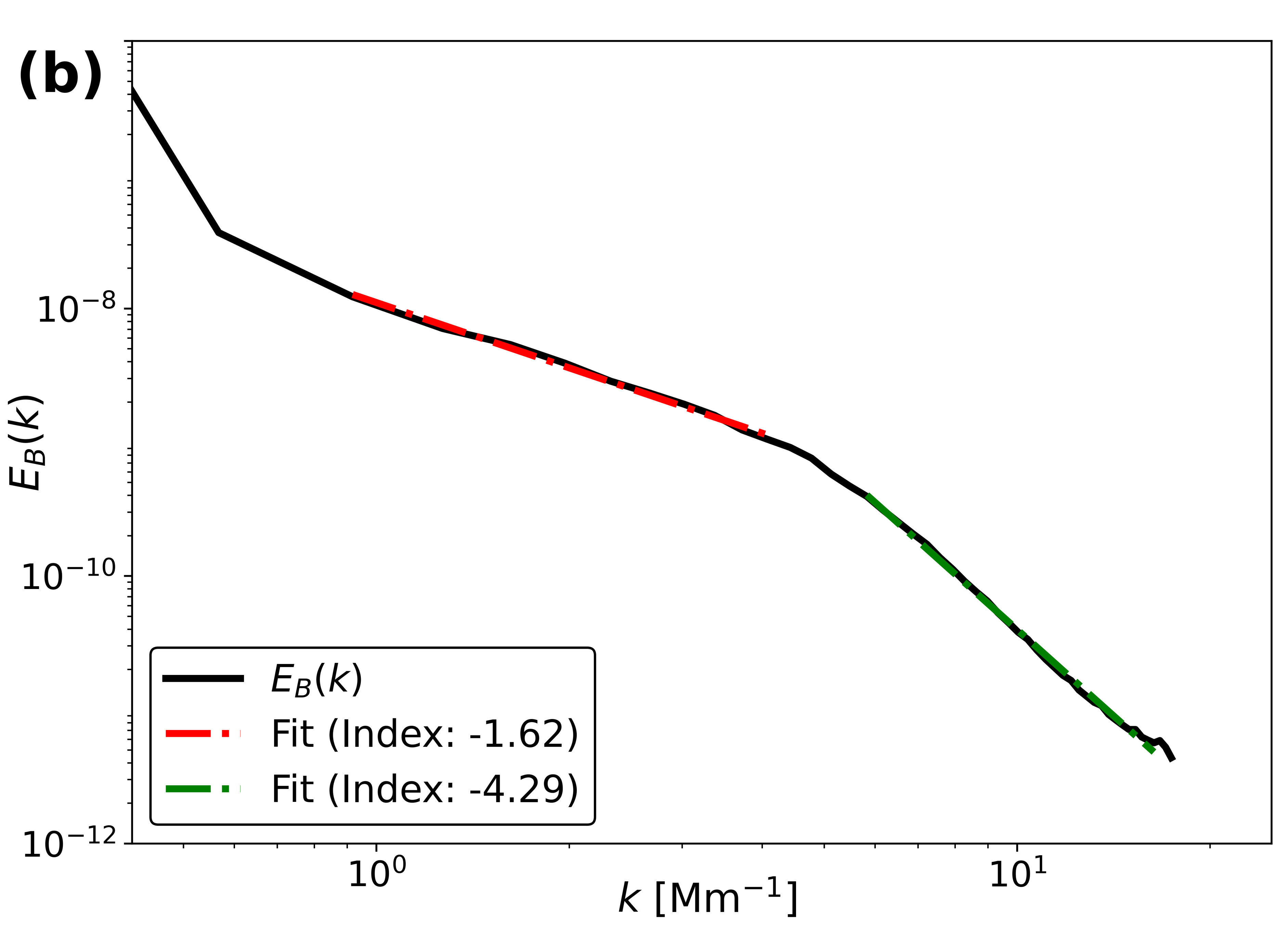}
    \caption{Panel (a): Distribution of $\Delta \mathrm{B}_i/\overline{\mathrm{B}}$, only accounting for the perturbation of grid points with strong $J_{\parallel}$ from region (\textbf{III}) in Figure \ref{fig:J_par_histogram}(a). The vertical dashed line marks the value 0.6. Panel (b): Magnetic energy density spectrum. Two distinct power-law regimes are evident: one at low wavenumbers with index $-1.62$ that is close to the Kolmogorov index, and another at high wavenumbers with a larger slope, $-4.29$. Both plots refer to $\mathrm{t} = 122.8$ min.}
    \label{fig:spectra}
\end{figure}

Since the ambient field provides a constant background, we can quantify the local magnetic field perturbation at each grid point relative to the background.  For each snapshot, we compute the global mean field $\overline{\mathrm{B}}$ by summing the magnitude of the magnetic field over all coronal grid points and dividing by the total number $\mathrm{N}$ of coronal grid points, $\overline{\mathrm{B}} = \sum_{i} \mathrm{B}_{i}/\mathrm{N}$. Then we determine the perturbation at each grid point as $\Delta \mathrm{B}_i/\overline{\mathrm{B}} = (\mathrm{B}_i - \overline{\mathrm{B}})/\overline{\mathrm{B}}$. For $\mathrm{t} = 122.8$ min, we show the distribution of the values of $\Delta \mathrm{B}_i/\overline{\mathrm{B}}$ in Figure \ref{fig:spectra}(a). The distribution exhibits a triple power-law, with the power-law at the highest fluctuations lying in the range $\Delta \mathrm{B}_i/\overline{\mathrm{B}} > 0.6$. 15\% of the perturbations fall into this range, and the range reaches 3. This clearly indicates strong turbulence in the system.

To gain deeper insight into how magnetic energy is distributed across different spatial scales, we conduct a spectral analysis restricted to the corona. Due to the open boundary conditions in both the x and z directions, as well as the isolation of the coronal sub-volume, the domain is not periodic. Therefore, we apply a Hann window function along both the $x$ and $z$ axes. We then compute the discrete Fourier transform of the magnetic field components. From these transformed components, we calculate the 3D spectral magnetic energy density $E_{B}(k)$, at every point in $k$-space using,
$
    E_{B}(k) = \big{(} |B_{x}(k)|^{2} + |B_{y}(k)|^{2} + |B_{z}(k)|^{2} \big{)}/2.$
In order to derive a one-dimensional averaged spectrum, we divide the $k$-space into bins according to the wave-vector magnitude $|k| = \sqrt{k_{x}^{2} + k_{y}^{2} + k_{z}^{2}}$, and we sum the spectral energy densities within each bin.

We focus on the later phase of the simulation, when many structures are already present and a turbulent regime has developed. Figure \ref{fig:spectra}(b) shows the magnetic energy spectra in the corona at $t = 122.8$ min. Small wavenumbers correspond to large spatial scales, and in this region the magnetic energy density spectrum has a power-law slope of $\alpha = -1.62$, close to the Kolmogorov index of $-5/3$. Based on the results shown in Figure \ref{fig:J_par_3D}, we estimate that the large-scale current sheets reach sizes of roughly 10--20 Mm before they begin to fragment. At higher wavenumbers, associated with smaller structures, the spectrum becomes significantly steeper, with a slope of $\alpha = -4.29$. This steeper slope further underscores the importance of reconnection-mediated magnetic structures (see the discussion in \cite{2008NPGeo..15...95A,Dong22}). In subsection \ref{cluster_analysis} below, we measure the characteristic sizes of the small-scale clusters of intense current sheets that form as the initial 10–20 Mm structure fragments (see region III in Figure \ref{fig:J_par_histogram}(a) and Figure \ref{fig:clusters_3D}), and we show that their typical scale is $<1$ Mm.

The results of this subsection suggest that the system resides in a highly turbulent regime, where the large-scale turbulence, in the 1--10 Mm range, exhibits Kolmogorov-like cascades, while the resulting small-scale structures with characteristic size $<$ 1 Mm trigger reconnection, transferring energy to particles through heating and acceleration.

\subsection{Fractal dimension analysis in three dimensions}

To characterize the spatial arrangement of the strongest energization sites, we compute the fractal dimension of the three-dimensional collection of locations where the parallel current exceeds a chosen threshold. For this purpose, we apply the standard box-counting technique \citep{Feder1988,Falconer2003}. In the box-counting method, the entire grid is covered by boxes of varying linear size $\ell$, and if the number of boxes $N(\ell)$ containing at least one supercritical grid point scales with $\ell$ as a power-law,
\begin{equation}
N(\ell)\propto \ell^{-D_f},
\end{equation}
then the power-law index $D_f$ is the fractal dimension.

\begin{figure}
    \centering
    \includegraphics[width=0.45\textwidth]{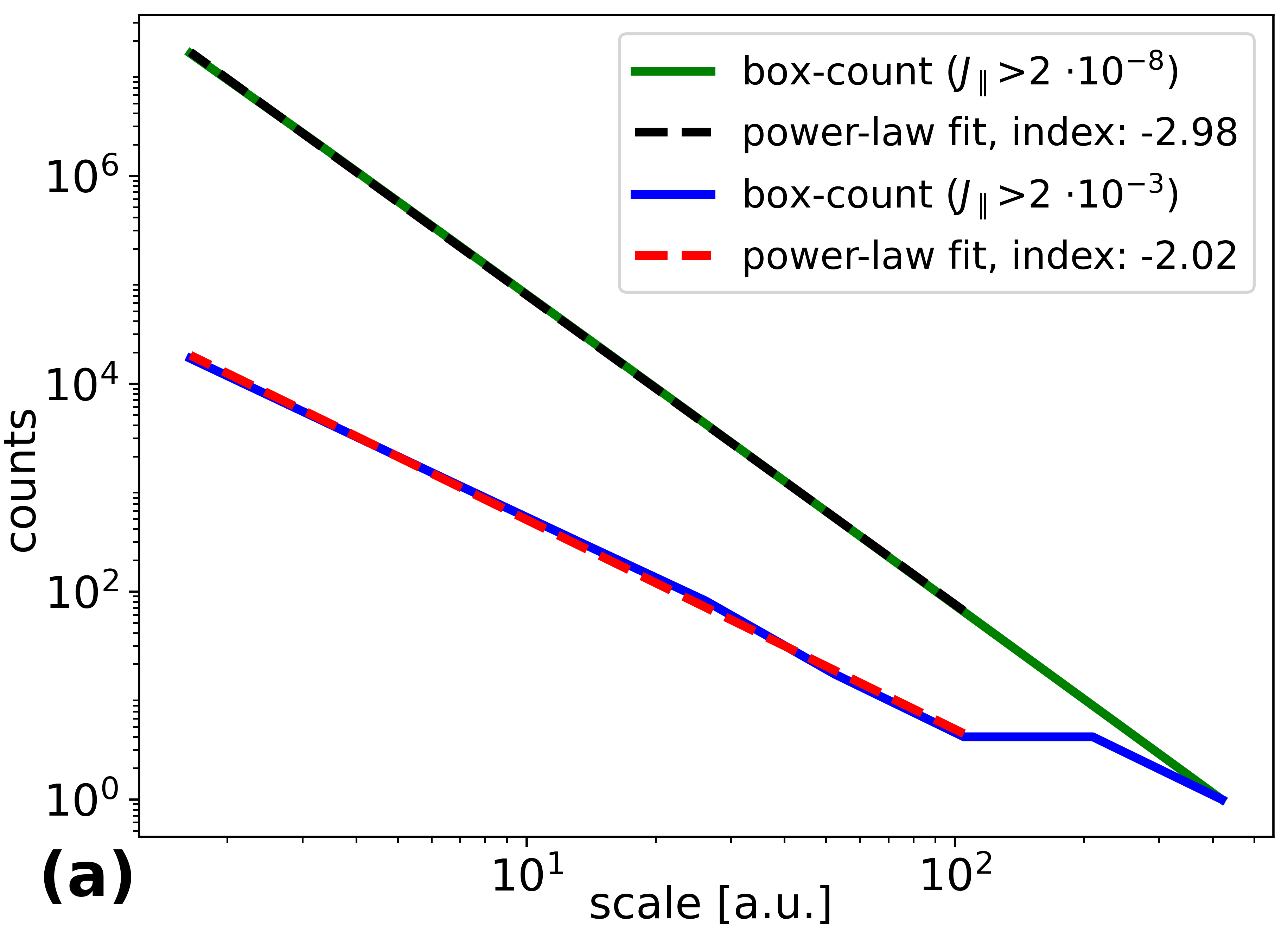}
    \includegraphics[width=0.45\textwidth]{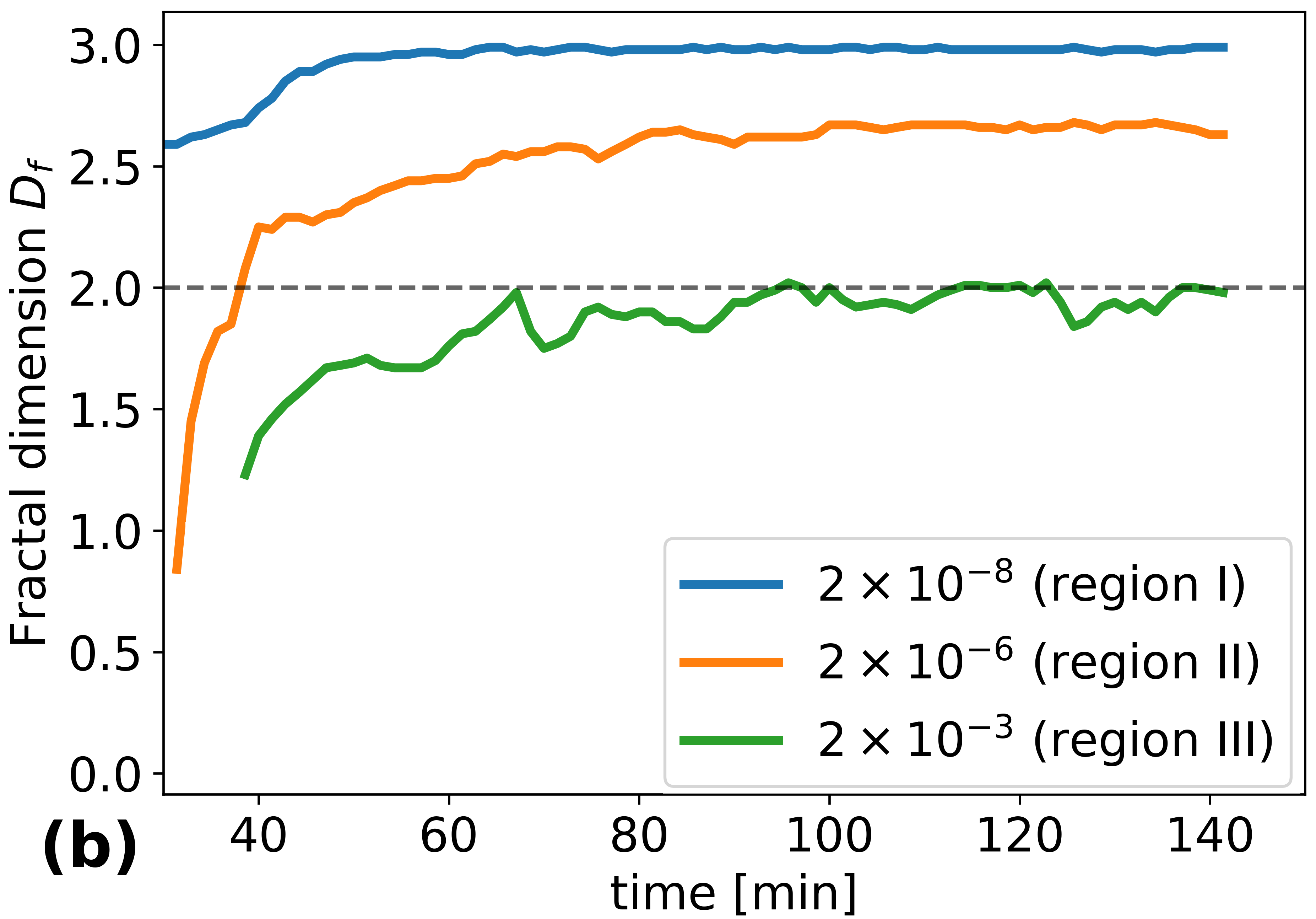}
    \caption{Panel (a) shows the calculation of the fractal dimension $D_{f}$ in the coronal region for time $t=122.8$ min for two different thresholds, one in region (\textbf{I}) and one in region (\textbf{III}) of Figure \ref{fig:J_par_histogram}(a). Panel (b) shows the time evolution of $D_{f}$ for the thresholds $J_{\parallel,th} = 2\cdot 10^{-8}$, $J_{\parallel,th} = 2\cdot 10^{-6}$, and $J_{\parallel,th} = 2\cdot 10^{-3}$ that fall into the three different regions of the parallel current distribution in Figure \ref{fig:J_par_histogram}(a). }
    \label{fig:fractal}
\end{figure}

We apply the analysis to successive stages of evolution, beginning with the formation of the standard jet and continuing into the blowout phase, when strong interaction develops between the erupting flux rope and the fragmented ambient field. As with the rest of the analysis, we show in panel (a) of Figure \ref{fig:fractal} the inferred fractal dimension at $t=122.8$ min and restrict the calculations in the corona. We have two estimates corresponding to different thresholds, one at $J_{\parallel,th} = 2 \, \cdot 10^{-3}$ and one at $J_{\parallel,th} = 2 \, \cdot 10^{-8}$, which fall into region (\textbf{III}) and (\textbf{I}) in Figure \ref{fig:J_par_histogram}(a), respectively. The power-law fit index for each curve provides the fractal dimension, which is $D_f = 2.02$ for $J_{\parallel} > 2 \, \cdot 10^{-3}$ and $D_f = 2.98$ for $J_{\parallel} > 2 \, \cdot 10^{-8}$. This implies that with the high threshold the structures are surface- or sheet-like, while with the low threshold they are space-filling. \citep{Feder1988,Falconer2003}. We also note that for the threshold $J_{\parallel} > 2  \cdot 10^{-3}$, there are very few structures of large size, as shown by the turn over of the scaling at $10^{2}$ in Figure \ref{fig:fractal}(a). The largest scales above $10^{2}$ have not been taken into account in the fractal dimension estimate.

We have also calculated the fractal dimension as a timeseries for the whole duration of the simulation, for the three different regions of $J_{\parallel}$ in Figure \ref{fig:J_par_histogram}(a). In Figure \ref{fig:fractal}(b), we show the evolution of the fractal dimension through time, for the Gaussian core (region \textbf{I}) with $J_{\parallel,th} = 2 \, \cdot 10^{-8}$, for the first power law tail (region \textbf{II}) with $J_{\parallel,th} = 2 \, \cdot 10^{-6}$ and for the second power law tail (region \textbf{III}) with $J_{\parallel,th} = 2 \, \cdot 10^{-3}$. We see that for currents from the Gaussian region, the fractal dimension remains close to $D_{f} \approx 3$, the weak currents thus are space-filling. The first power law is an intermediate region where the fractal dimensions start from $D_{f} \approx 1$ and quickly surpass $D_{f} \approx 2$, before maintaining a steady plateau at $D_{f} \approx 2.5$. The second power law region has the strongest currents, and while at the beginning it has a smaller fractal dimension, it quickly reaches and maintains $D_{f} \approx 2$ for the rest of the simulation. The values obtained here therefore indicate that the intense $J_{\parallel}$ structures are predominantly quasi-two-dimensional: they are spatially extended in two directions and remain thin in the third. 

This outcome admits a straightforward physical interpretation. It indicates that the turbulent, structured atmosphere is not homogeneously packed with dissipative regions; instead, it is threaded by a broken network of thin, current-sheet-like formations, in line with the expected intermittent morphology of nonlinear MHD evolution \citep{biskamp2003magnetohydrodynamic,Frisch_1995}. Equally significant is that $D_f$ remains nearly constant throughout the eruptive phase, implying that although the eruptions intensify the active structures and enhance their spatial complexity, they do not fundamentally modify their topological character. Thin, fragmented, sheet-like coherent structures continue to dominate the system’s dynamics over the entire evolution. This geometry is particularly crucial for turbulent heating and particle acceleration, as it implies that energy conversion is concentrated on localized surfaces—fragmented current sheets and reconnection layers—rather than being spread uniformly across the volume.

We also expect a current sheet to have high values of $J_{\parallel}$ while at the same time the magnetic field of the structure shouldn't be excessively large \citep{2021ApJ...914...71I,2026A&A...705A..86N}. The lower limit for identifying a potentially reconnecting current sheet is inversely proportional to the grid resolution, $J_{\parallel}/\mathrm{B} > 1/2d$ \citep{jiang2016data}, where $d$ is the grid length. In our simulation the ratio the theoretical minimum is $J_{\parallel}/\mathrm{B} \approx 10^{-6}$. For the strong current structures from region (\textbf{III}), we find a ratio of $J_{\parallel}/\mathrm{B} > 0.1$, which affirms that the strong currents of region (\textbf{III}) are associated with current sheets. 

\subsection{Cluster analysis and energetics}\label{cluster_analysis}

To characterize the spatial coherence of strong parallel current structures, we carry out a threshold-based cluster analysis. In particular, we consider all grid points that satisfy $J_{\parallel} > 2 \times10^{-3},$ using the highest threshold from the fractal analysis above, to isolate the high-valued tail of the $J_{\parallel}$ distribution (region III in Figure \ref{fig:J_par_histogram}a). Starting from any point that exceeds the threshold, the algorithm searches for neighboring points that meet the same criterion and groups all mutually connected points into a single cluster, where the neighborhood is defined as the six nearest grid points on the 3D Cartesian grid. This recursive process continues until no further above-threshold neighbors can be incorporated. The search then restarts from any other unassigned point above the threshold. In this manner, above-threshold structures that are spatially close yet not first-order topologically linked are identified as separate clusters.

\begin{figure}[htp!]
    \centering
    \includegraphics[width=0.8\textwidth]{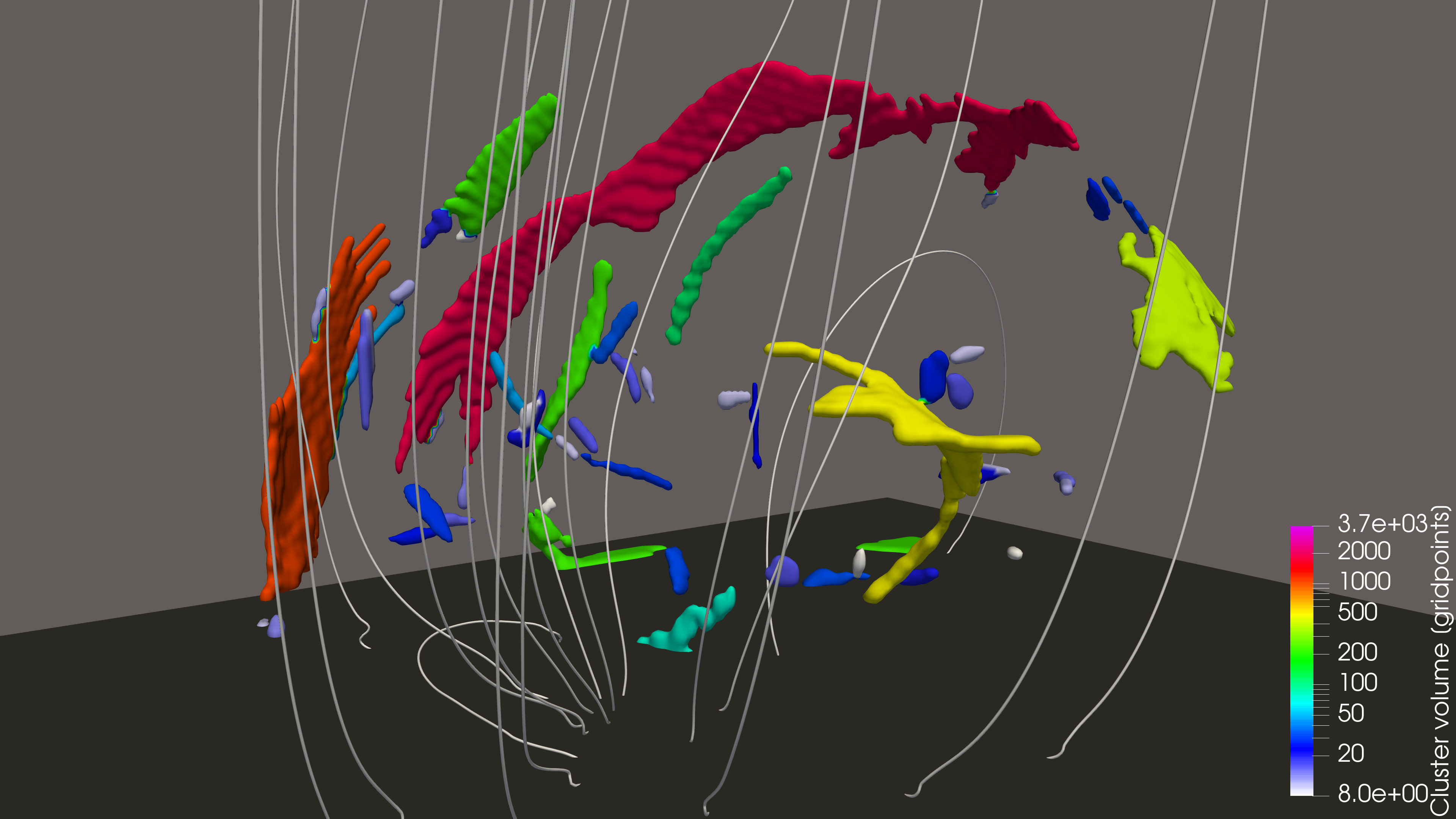}
    \caption{3D view of all clusters with $J_{\parallel} > 0.002$ [A m$^{-2}$] at $\mathrm{t}=122.8$ min. The clusters are colored according to their volume and field lines depict the magnetic field structure in the vicinity. The evolution of the clusters can be seen in the \href{https://drive.google.com/file/d/1eWTXFmrf_vc-0j4l6PgrBKws4DbxJZLk/view?usp=sharing}{\color{red}animation} of this figure, with many clusters of various sizes being present at every timeframe.}
    \label{fig:clusters_3D}
\end{figure}

\begin{figure}[htp!]
    \centering
    \includegraphics[width=0.45\textwidth]{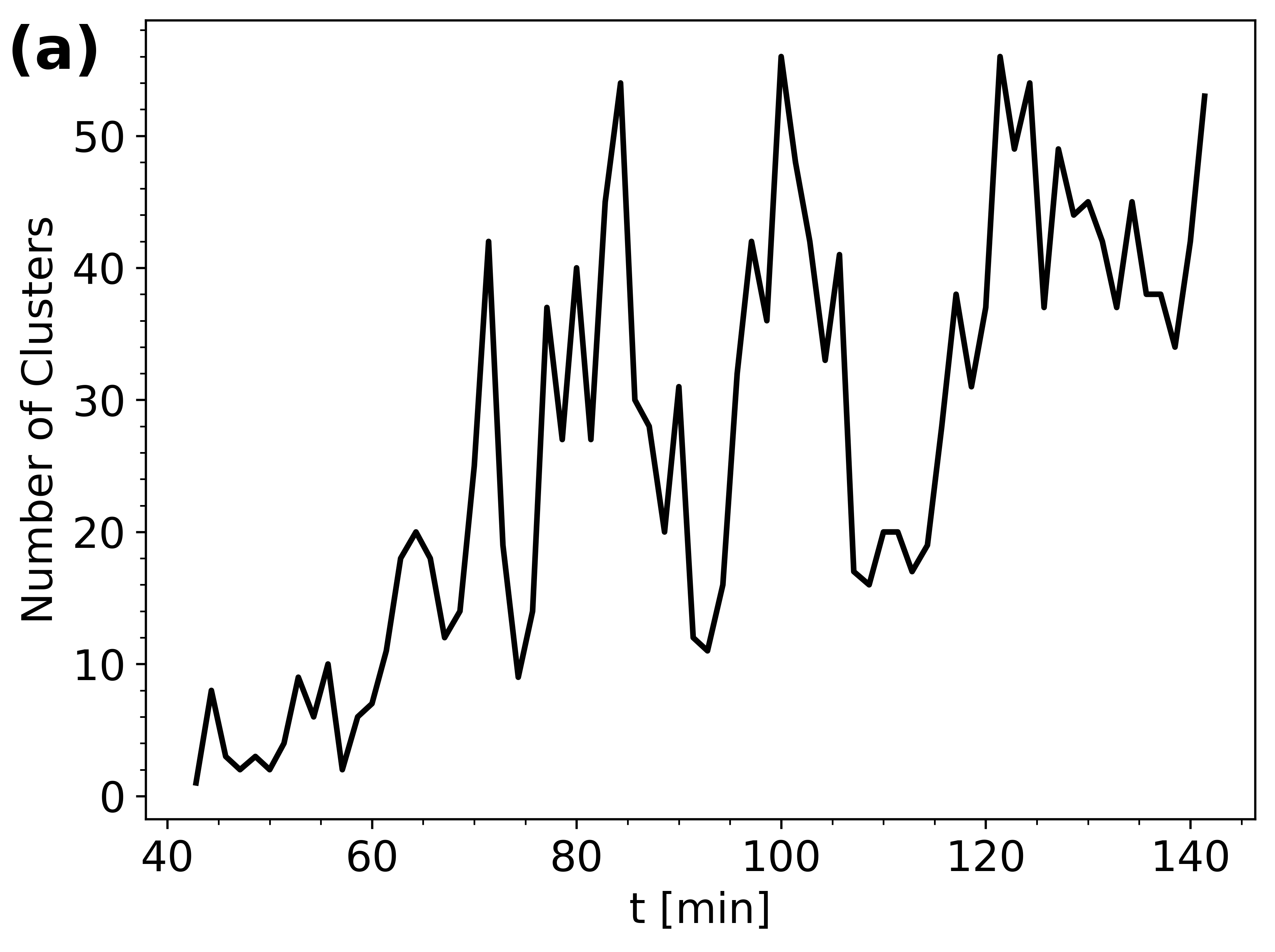}
    \includegraphics[width=0.45\textwidth]{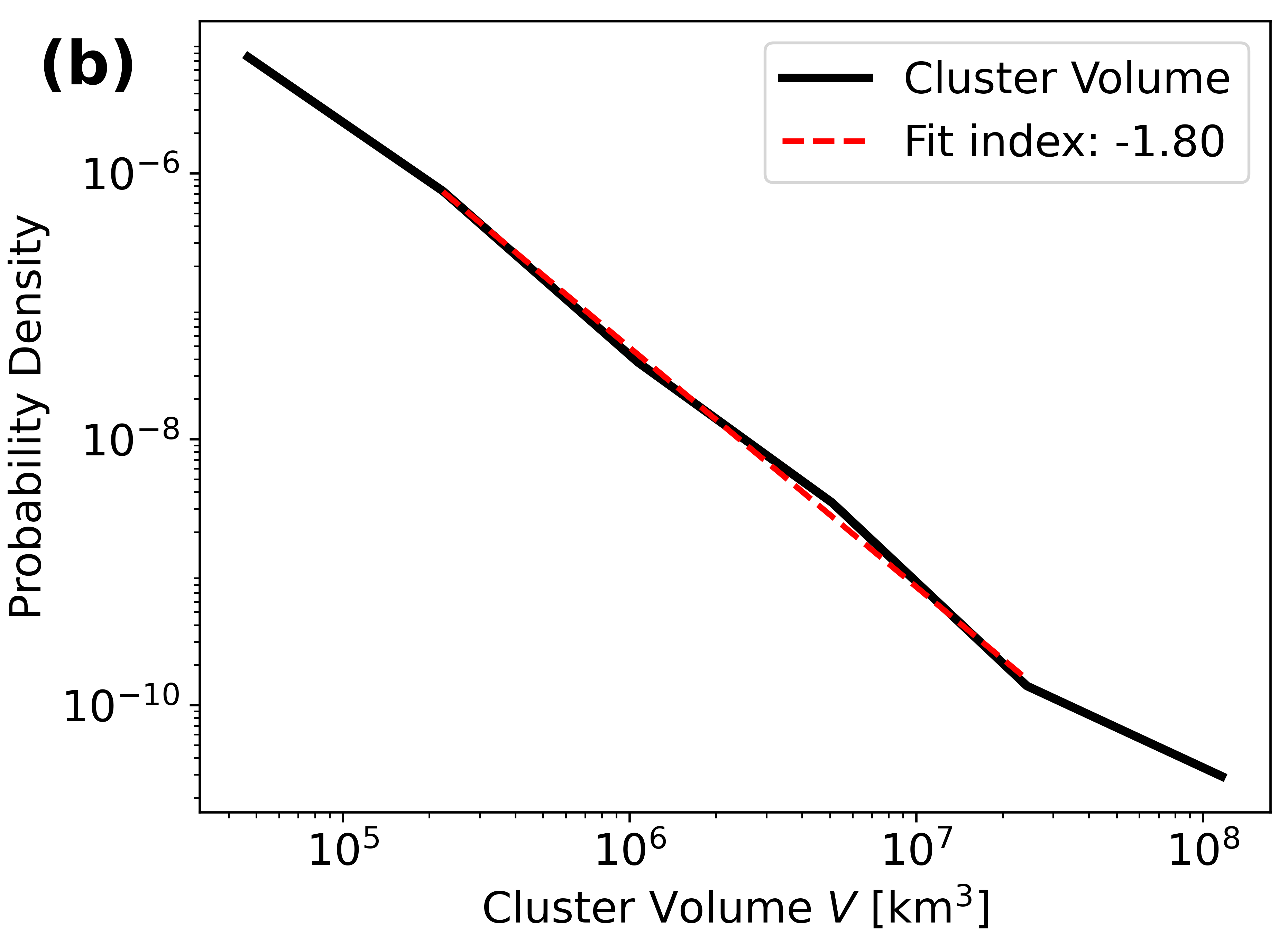}
    \caption{Panel (a): Timeseries of the total number of clusters that has an increasing trend. Panel (b) shows the distribution of the cluster volume at $\mathrm{t}=122.8$ min, forming a power law for the intermediate and small clusters, and a different power law for the large clusters.}
    \label{fig:cluster_volume}
\end{figure}

To visualize the clusters, we plot all points belonging to a cluster, in Figure \ref{fig:clusters_3D} at $\mathrm{t} = 122.8$ min where four additional flux tubes have erupted, with colors indicating the cluster sizes. We observe that after the fourth blowout jet, many clusters of various sizes are present at the boundary between open and closed field lines. The full evolution of the clusters can be seen in the \href{https://drive.google.com/file/d/1eWTXFmrf_vc-0j4l6PgrBKws4DbxJZLk/view?usp=sharing}{\color{red}animation} of Figure \ref{fig:clusters_3D}, where at every time frame there are many clusters of various sizes. The absolute number of clusters varies with time and their total count has an increasing trend, as illustrated in Figure \ref{fig:cluster_volume}(a). We observe an increasing trend amid large fluctuations for approximately 100 minutes, despite four major eruptive events that expel material upward during this timeframe. For the statistical analysis of the clusters, we require that a cluster contain at least 10 gridpoints and up to $10^{6}$ gridpoints, thereby excluding small, highly localized above-threshold spots and the single large cluster representing the new flux rope forming in the lower corona.  

Panel (b) of Figure \ref{fig:cluster_volume} shows the resulting distribution of cluster volumes at $t=122.8$ min. The distribution is characterized by a power-law regime, with an index $\alpha = -1.8$ for the intermediate and small clusters; the vast majority of clusters are small, spanning only a few grid points, while only a relatively small fraction reach volumes on the order of $10^{8}$ km$^{3}$.

The clusters exhibit an average characteristic Alfv\'en timescale, $\tau_{Aj} \sim L_j/V_{Aj}$ (s), estimated from the cluster size $L_j$ and the local Alfv\'en speed $V_{Aj}$. For representative parameters ($L_j \approx 1-10$ Mm, $B \approx 100$ G, and $n \approx 10^9\,\mathrm{cm^{-3}}$), the inferred cluster Alfv\'en timescale is of the order of $\sim 1$ s. The clusters are observed to disintegrate over timescales spanning over \(\sim 10-10^{2}\) Alfv\'en times, which is consistent with the characteristic timescale for current-sheet fragmentation reported by \citet{Onofri06} and \citet{WangYulei25}.

\begin{figure}[htp!]
    \centering
    \includegraphics[width=0.45\textwidth]{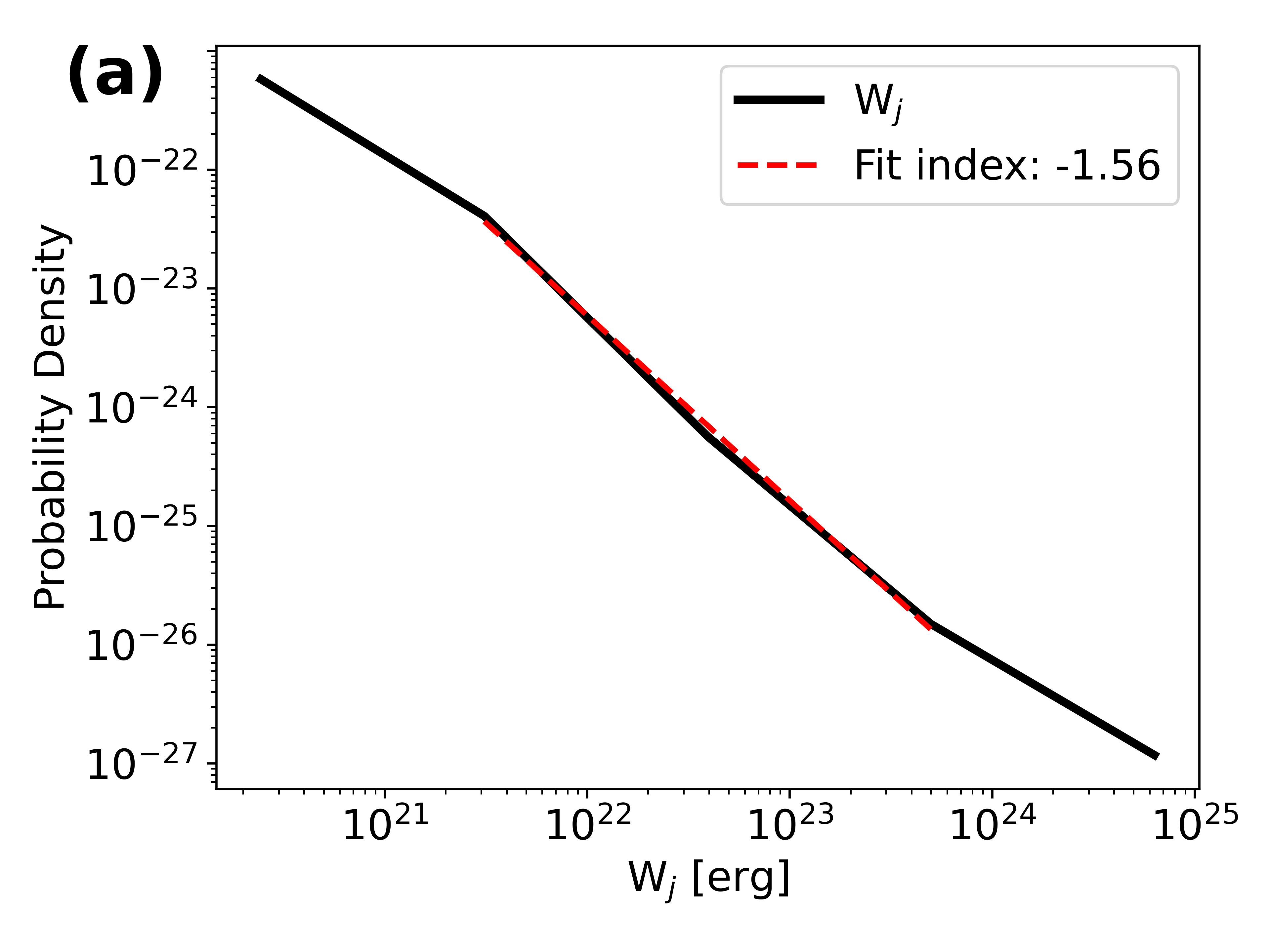}
    \includegraphics[width=0.45\textwidth]{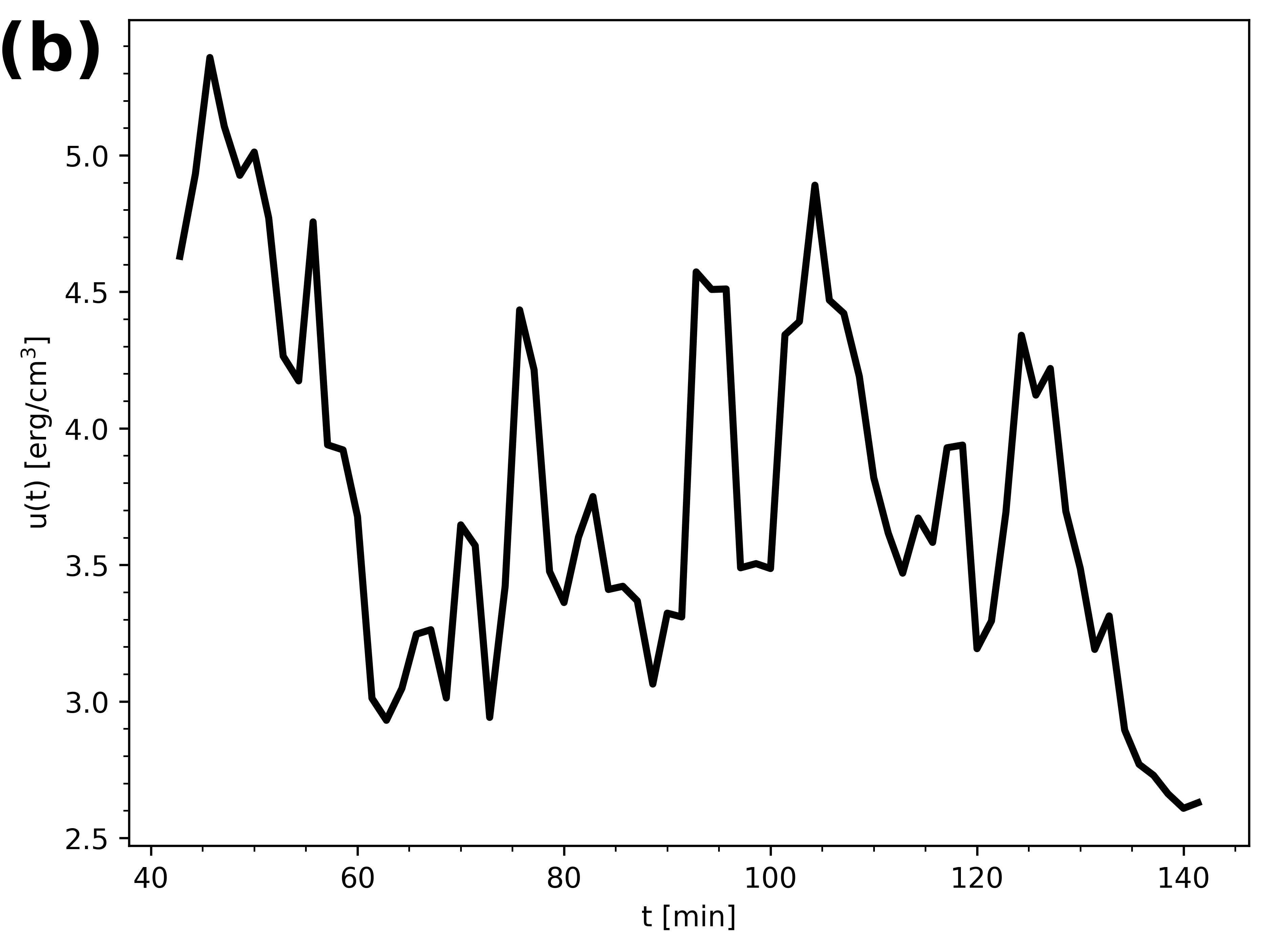}
    \caption{Panel (a): Energy distribution of clusters with strong parallel current at $\mathrm{t}=122.8$ min. Panel (b): Time series of total energy that is dissipated from those clusters, normalized with their volumes.}
    \label{fig:energetics}
\end{figure}

Beyond the cluster physical properties described in the previous section, we now investigate how much energy these structures can dissipate. We adopt the same criterion to detect clusters, namely $J_{\parallel,th} = 0.002$ [A $m^{-2}$]. For each identified cluster, we add up the values of $J_{\parallel}$ over all grid points that belong to it. To evaluate the Ohmic heating produced by an individual cluster, we estimate its released energy as \[\mathrm{W_j} = ( \sum_{i} \mathrm{\eta} \ J_{\parallel,i}^{2} \ V_{i})\ \mathrm{\delta t}_{j},\] where $V_{i}$ denotes the volume of a grid cell, $J_{\parallel,i}$ is the parallel current density at that cell, and $\mathrm{\delta t}_{j}$ is the time span over which the cluster persists. Each cluster has a different lifetime, based on its Alfv\'en time, which we estimate here as $\mathrm{\delta t}_{j} \sim 100 \cdot \tau_{Aj}$ seconds, particularly for the intermediate and larger clusters (see the discussion just above). Figure \ref{fig:energetics}(a) shows the probability distribution of cluster energy at $t=122.8$ min. The distribution displays a clear power-law trend with index $\alpha = -1.56$, and cluster energies spanning from $10^{20}$ to $10^{25}$ erg.

We also calculated the total resistive energy deposited by the intense currents within the entire set of clusters at a given time, normalized by the aggregate volume of all clusters, expressed as
\begin{equation}
    \mathrm{u}(\mathrm{t}) = \sum_{j} \frac{W_{j}}{V_{J}}
\end{equation}
The sum is over all grid points of all clusters at a given timestep. Plotted as a function of time in Figure \ref{fig:energetics}(b), this quantity illustrates the temporal evolution of the energy dissipation during the simulation. After the initial fragmentation of the standard jet’s current sheet, the energy dissipation is strongest; it then settles to slightly lower values during the blowout jets. At later times, fewer current sheets undergo reconnection; consequently, the total released energy decreases.  

Neglecting energy losses, an injected energy density in the range of $u = 3\text{--}5 \text{ ergs cm}^{-3}$ (see panel (b) of Figure \ref{fig:energetics}) leads to a local temperature rise of $\Delta T \sim u / (3nk_B)\sim 10 \text{ MK}$, assuming the ideal gas law and a representative coronal number density of $n = 10^9 \text{ cm}^{-3}$, where $k_B$ is the Boltzmann constant. In reality, radiative and conductive cooling will counteract the resistive heating and lower this temperature increase, but a detailed treatment of these loss processes and the formation of extended hot structures lies beyond the scope of this study.

\section{Discussion and Summary}\label{summary}

 We investigated how the 3D large-scale current system that develops during the emergence of magnetic flux tube becomes disrupted and transforms into a multiscale ensemble of coherent current structures. To characterize this transformation, we examined the statistical, spectral, geometrical, and energetic properties of the current structures, focusing in particular on the parallel current. The key findings are summarized below.

\begin{enumerate}
	\item The fragmentation of current sheets formed at the interface between, emerging magnetic flux tubes and weak, open magnetic field lines leads to the development of numerous intense current sheets within the closed field lines region, along with comparatively weak, filamentary currents aligned with the open field lines. In this article, our analysis focuses on the strong current structures that arise within closed magnetic topologies.
    \item The probability distribution of the parallel current in the closed magnetic field lines region is strongly non-Gaussian. It consists of a Gaussian-like core at low values and two extended power-law tails at higher values.  Only a small fraction of the current population, approximately $1\%$, belongs to the strongest current regime. This small population is nevertheless physically important because it identifies the most intense, spatially intermittent current structures, where reconnection and enhanced dissipation are most likely.

    \item The magnetic energy spectrum shows two distinct scaling ranges. At large spatial scales, between 1-10Mm the spectrum approaches a power-law with a Kolmogorov-like slope. At smaller scales, below 1 Mm, the spectrum steepens substantially in the later stages of the simulation. This spectral break indicates that the system is not characterized by a single inertial-range cascade. 

    \item The three-dimensional box-counting analysis shows that the most intense current structures have a fractal dimension close to $D_f \simeq 2$. This result holds throughout the later eruptive phase and indicates that the dominant dissipative structures are sheet-like rather than filamentary or volume-filling. Weaker current populations have larger fractal dimensions and occupy a much larger fraction of the coronal volume.

    \item The cluster analysis confirms the intermittent and multiscale nature of the fragmented current system. The cluster volume distribution follows a power law, with many small clusters and far fewer large structures.  This hierarchy of cluster sizes shows that disrupting the original current sheet produces a multiscale, spatially intermittent population of coherent structures rather than a random collection of isolated current concentrations. Current sheets are intermittent structures, and their lifetime is tied to their characteristic Alfvén time. The estimated Ohmic energy released by the clusters also follows a power-law distribution with characteristic slope 1.56, extending from approximately $10^{20}$ to $10^{25}$ erg. The energy linked to each structure is strongly tied to its characteristic length, suggesting that larger coherent structures prevail at the high-energy end of the distribution. However, the total energy density released by the strongest-current clusters increases slightly after the initial eruption, and it then decreases during the later stages of the evolution. 
\end{enumerate}

The small filling factor of the strongest current structures is particularly important. Although these structures occupy only a minor fraction of the volume, they provide the most favorable sites for impulsive energy release. This is consistent with a picture in which coronal heating and particle acceleration are spatially localized, temporally intermittent, and controlled by the statistics of coherent structures rather than by volume-averaged plasma properties.

A further caveat concerns our use of a fixed, relatively large uniform resistivity, which was chosen for numerical stability rather than to match the (much higher) Lundquist numbers expected in the solar corona.  We therefore regard the qualitative picture established here — a single global current sheet self-consistently fragmenting into a hierarchy of sheet-like, power-law-distributed coherent structures — as robust, while the precise quantitative indices should be treated as representative of this resolution and resistivity rather than as final, resolution-converged values. A systematic resolution/resistivity scan is left for future work.

The present study is based on the MHD evolution of the eruptive flux-emergence system and therefore does not directly follow the kinetic response of particles to the fragmented electromagnetic fields.  A natural next step is to use the time-dependent electric and magnetic fields from these simulations as input for test-particle simulations to determine how the evolving hierarchy of coherent structures heats and accelerates particles during recurrent eruptive activity.

In conclusion, our findings demonstrate, for the first time, that the repeated emergence of eruptive flux converts an initially large-scale current sheet—created during the initial magnetic field emergence and subsequent eruptions—into a fragmented, multiscale, turbulent region over several hundred Alfvén times. The most intense dissipative structures are rare, sheet-like, and spatially intermittent, and their volumes and energies exhibit power-law distributions.

\begin{acknowledgments}
We thank the referee for their significant and insightful feedback. We would also like to thank Vera Agalianou and Juxhin Zhuleku for their important contributions to this work. The numerical simulation was performed on the ARIS HPC of the National Infrastructures for Research and Technology S.A. (GRNET S.A.) under the project name: Emergence. This work has been supported by the European Research Council through the Synergy Grant No.810218 (“The Whole Sun”, ERC-2018-SyG).
\end{acknowledgments}

\bibliographystyle{aasjournalv7}
\bibliography{refs_apj}

\end{document}